\documentclass[acmsmall]{acmart}

\AtBeginDocument{%
  }

\begin{document}

\title{Adversarial Patterns: Building Robust Android Malware Classifiers}


\author{Dipkamal Bhusal} 
\email{db1702@rit.edu}
\affiliation{%
  \institution{Department of Software Engineering, Rochester Institute of Technology}
  \streetaddress{134 Lomb Memorial Dr}
  \city{Rochester}
  \state{NY}
  \country{USA}
}

\author{Nidhi Rastogi} 
\email{nxrvse@rit.edu}
\affiliation{%
  \institution{Department of Software Engineering, Rochester Institute of Technology}
  \streetaddress{134 Lomb Memorial Dr}
  \city{Rochester}
  \state{NY}
  \country{USA}
}








\renewcommand{\shortauthors}{Bhusal et al.}

\begin{abstract}
  Machine learning models are increasingly being adopted across various fields, such as medicine, business, autonomous vehicles, and cybersecurity, to analyze vast amounts of data, detect patterns, and make predictions or recommendations. In the field of cybersecurity, these models have made significant improvements in malware detection. However, despite their ability to understand complex patterns from unstructured data, these models are susceptible to adversarial attacks that perform slight modifications in malware samples, leading to misclassification from malignant to benign. Numerous defense approaches have been proposed to either detect such adversarial attacks or improve model robustness. These approaches have resulted in a multitude of attack and defense techniques and the emergence of a field known as `adversarial machine learning.' In this survey paper, we provide a comprehensive review of adversarial machine learning in the context of Android malware classifiers. Android is the most widely used operating system globally and is an easy target for malicious agents. The paper first presents an extensive background on Android malware classifiers, followed by an examination of the latest advancements in adversarial attacks and defenses. Finally, the paper provides guidelines for designing robust malware classifiers and outlines research directions for the future.
\end{abstract}




\maketitle

\section{Introduction}

Android operating system has one of the largest smartphone market shares and is one of the most targeted platforms by malware attackers \cite{malwarestatistics}. Researchers at the cybersecurity company Kaspersky have identified a concerning trend in past years. Applications, called apps, downloaded from the Google Play Store, the primary app store for Android-compatible mobile devices, were malicious \cite{bewaremalicious}. A longitudinal study further revealed that anti-malware solutions detected around 5 million malicious apps on mobile platforms in the first quarter of 2023 alone \cite{malwareevolution}. In response, cybersecurity experts aim to subvert these attacks to keep mobile devices safe to use. They accomplish this by building malware detection systems that identify malicious apps and reject them prior to installation. Machine learning models are predominantly used to build these classifiers. These models have replaced signature-based heuristic methods and have reduced dependence on handcrafted features for malware classification, thereby assuring significant improvements in accuracy and efficiency \cite{arp2014drebin}.

\begin{figure}[]
\centering
\includegraphics[width=0.60\textwidth]{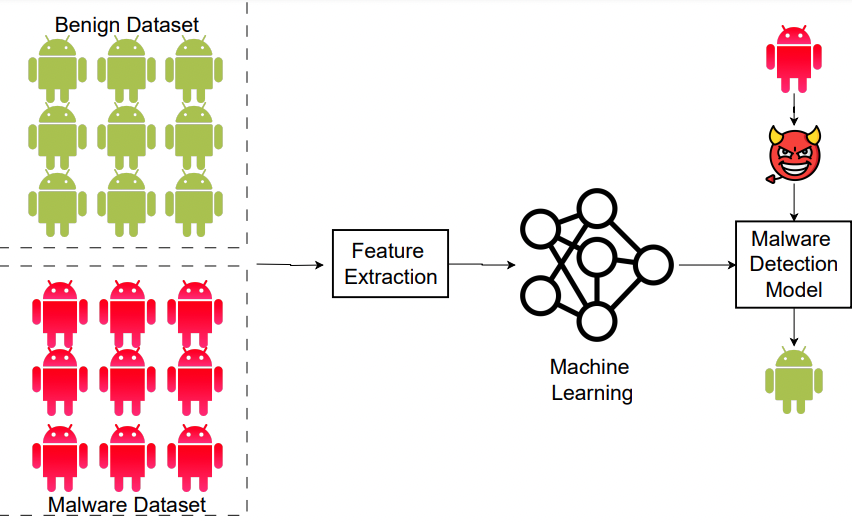}
\caption{A conceptual representation of evasion attack in a malware classifier where a test sample of malicious nature is misclassified as benign.}
\label{fig_evasion_malware}
\end{figure}

While ML-based malware classifiers perform more competently with identifying inputs more accurately than even a cybersecurity expert would classify, they perform poorly to subtle yet maliciously crafted data. Malware classifiers are vulnerable to such perturbation, known as adversarial samples \cite{szegedyintriguing}, but continue to classify Android applications with very high confidence, albeit with unexpected results. An adversary can model these manipulations during the training process or when testing the model. When implementing an evasion attack, an adversary crafts a test sample to evade the original classification decision. For example, carefully manipulated code snippets are placed so that a model classifies a malware sample as benign. Figure \ref{fig_evasion_malware} shows a conceptual representation of an evasion attack in an Android malware classifier. In poisoning attacks, fake training data can be injected into the training dataset, inducing erroneous predictions in these classifiers. State-of-the-art classifiers have been shown to be vulnerable against both kinds of attacks \cite{szegedyintriguing,goodfellow2015explaining,kurakin2016adversarial,papernot2016limitations,moosavi2016deepfool, papernot2017practical, carlini2017towards, grosse2017adversarial, yang2017malware, chen2019android, pierazzi2020intriguing}.

\begin{figure}[]
\centering
\includegraphics[width=0.75\textwidth]{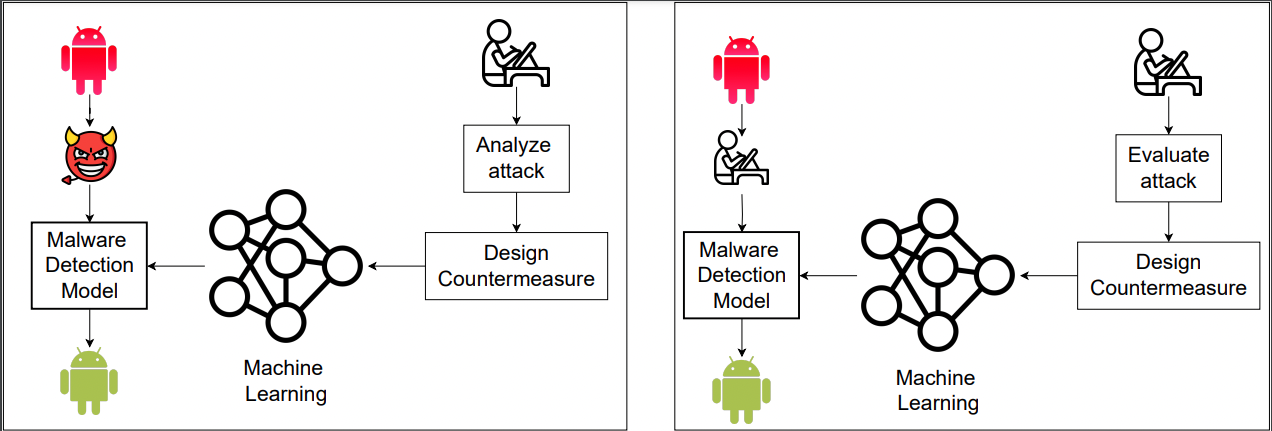}
\caption{A schematic representation of two types of adversarial defense. Left: reactive, Right: proactive.}
\label{fig_defense}
\end{figure}

In contrast, countermeasures against adversarial attacks have been proposed to improve the security of malware classifiers. One approach, called proactive defense, incorporates the knowledge of the attacker before an actual attack. Cybersecurity experts design defense methods to \textit{robustify} the classifier or detect adversarial samples in order to mitigate the impact of the attack or prevent potential attacks. The second approach, which is most common in the real world, is called reactive defense, where machine learning models are modified after observing a new attack, which often involves retraining the classifier with new data or modifying features \cite{biggio2018wild}. However, this does not prevent an adversary from designing new attack algorithms. An adversary can exploit the limitation of these defenses by continually crafting new attack algorithms and evading the classifier. As a result, an arms race has started in adversarial malware detection. Figure \ref{fig_defense} shows a schematic representation of two types of defense.

\par
This paper is a comprehensive survey of the most significant research performed on evasion attacks and defenses for Android malware classifiers. We also provide all the necessary theoretical foundations that underpin these attacks and defenses. Prior work \cite{demetrio2021adversarial,li2021arms} is limited to Windows OS-based devices and PDF files and only briefly discusses Android mobile malware attacks. Grosse et al. \cite{grosse2023machine} analyze attack occurrence and concerns in real-world settings and report quantitative results. Aryal et al. \cite{aryal2021survey} present a survey on adversarial machine learning for malware analysis but only focus on attacks. Berger et al. \cite{berger2022problem} only provide a survey on problem-space evasion attacks in the Android operating system. However, given the current attention to Android malware, there is a need for a deeper and more comprehensive understanding. The main contributions of this paper are summarized below:
\begin{enumerate}
    \item This is the first survey that succinctly summarizes and critically evaluates the current body of research on both \textit{evasion attacks and defenses} on Android malware classifiers.
    \item We clarify the underlying principles and theoretical foundations behind the vulnerabilities of classifiers to adversarial samples, helping to build a more robust understanding of the domain.
    \item For researchers interested in exploring adversarial machine learning, we also provide essential evaluation criteria, guidelines, and future directions for its application to malware classifiers.
\end{enumerate}


\section{Relevant Background}\label{sec:background}
 In this section, we first briefly summarize relevant machine learning algorithms (\S \ref{sec:machinelearning}), then discuss the taxonomy of adversarial attack (\S \ref{sec:taxonomyofadversarial}). We introduce the necessary background on Android in \S \ref{sec:androidfile}, discuss malware classifiers (\S \ref{sec:malwareanalysis}) and finally present an example of an evasion attack (\S \ref{sec:exampleonandroid}).

\subsection{Machine Learning Models for Malware Classification}\label{sec:machinelearning}

\begin{table}[]
\centering
\caption{Overview of machine learning algorithms used in Android malware classifier}
\label{tab:my-table-ml}
\resizebox{\textwidth}{!}{%
\begin{tabular}{@{}lllll@{}}
\toprule
\multicolumn{1}{c}{\textbf{Algorithm}} &
  \multicolumn{1}{c}{\textbf{Key Characteristics}} &
  \multicolumn{1}{c}{\textbf{Pros}} &
  \multicolumn{1}{c}{\textbf{Cons}} &
  \multicolumn{1}{c}{\textbf{Related work}} \\ \midrule
k-NN &
  \begin{tabular}[c]{@{}l@{}}Instance based\\ Non-parametric\end{tabular} &
  \begin{tabular}[c]{@{}l@{}}Simple\\ No-training phase\end{tabular} &
  \begin{tabular}[c]{@{}l@{}}Computationally \\ expensive\end{tabular} &
  \begin{tabular}[c]{@{}l@{}}\cite{saracino2016madam, arora2018hybrid, kakavand2018application} \\ \cite{taheri2020similarity, dehkordy2021new}\end{tabular} \\ \midrule
SVM &
  Linear/Non-linear separation &
  Effective in high-dimension &
  Sensitive to noise &
  \begin{tabular}[c]{@{}l@{}}\cite{arp2014drebin, su2016deep,lindorfer2015marvin} \\ \cite{wang2014exploring, han2020enhanced, qiao2016merging}\\ 
  \cite{onwuzurike2019mamadroid, han2020enhanced}\end{tabular} \\ \midrule 
Naive Bayes &
  Probabilistic, Bayes’ theorem &
  Fast, Handles irrelevant feature &
  Assumes feature independence &
  \begin{tabular}[c]{@{}l@{}}\cite{li2018android, chen2018tinydroid, kang2016n}\\ \cite{xiao2016identifying, zhang2018novel}\end{tabular} \\ \midrule 
Decision Tree &
  Tree-like model, Rule-based &
  Easy to interpret &
  Prone to overfitting &
  \begin{tabular}[c]{@{}l@{}}\cite{wang2014exploring, wang2016trafficav, utku2018decision}\\ \cite{zulkifli2018android, hossain2020optimized}\end{tabular} \\ \midrule 
Random Forest &
  Ensemble of decision trees &
  Reduces overfitting &
  Computationally expensive &
  \begin{tabular}[c]{@{}l@{}}\cite{wang2014exploring, qiao2016merging, vinod2019machine}\\ \cite{thangavelooa2020datdroid,onwuzurike2019mamadroid}\end{tabular} \\ \midrule 
Neural Network &
  Deep learning model &
  Complex feature learning &
  \begin{tabular}[c]{@{}l@{}}Requires large datasets\\ Prone to overfitting\end{tabular} &
  \begin{tabular}[c]{@{}l@{}}\cite{hou2016deep4maldroid, li2018significant,kim2018multimodal} \\ \cite{chen2018tinydroid, bai2020famd, qiao2016merging}\end{tabular} \\ \midrule 
DBN &
  Deep probabilistic model &
  Captures complex patterns &
  Training complexity &
  \begin{tabular}[c]{@{}l@{}}\cite{su2016deep, yuan2016droiddetector, chen2019droidvecdeep} \\ \cite{hou2016droiddelver}\end{tabular} \\ \midrule 
LSTM/RNN &
  Recurrent memory connections &
  Handles sequential data &
  \begin{tabular}[c]{@{}l@{}}Vanishing/exploding gradient\\ Complex architecture\end{tabular} &
  \begin{tabular}[c]{@{}l@{}}\cite{xiao2019android, vinayakumar2018detecting, xiao2019android} \\ \cite{amin2020static}\end{tabular} \\ \midrule 
VAE &
  Variational inference &
  Continuous latent space &
  Training complexity &
  \cite{li2021robust, yumlembam2022iot, hou2021disentangled}\\ \midrule 
GAN &
  Generative adversarial network &
  Produces realistic data &
  Training instability &
  \cite{amin2022android, chen2021using, zhang2021enhanced} \\ \midrule 
Genetic Algorithm &
  Evolutionary algorithms &
  Global optimization &
  Convergence speed &
  \cite{vivekanandam2021design,fatima2019android,xie2023ga} \\ \bottomrule
\end{tabular}%
}
\end{table}

Machine learning models can detect new, unseen malware through the understanding of underlying behavioral patterns or structural features in malware applications. In this section, we discuss several such algorithms used for malware classification. Refer to Table \ref{tab:my-table-ml} for an overview of machine learning models and selected relevant works. 

\textbf{K-nearest Neighbor (KNN)}: KNN is a non-parametric and instance-based learning method that doesn't assume the distribution of the training data \cite{knn1}. It operates by identifying the K closest instances or data points in the training data to a given input and then assigns the class label based on the majority of the K neighbors.

\textbf{Support Vector Machine (SVM)}: SVM works by finding the hyperplane that maximizes the margin between classes in a high-dimensional feature space \cite{hearst1998support}. SVMs are especially effective in scenarios where the decision boundary is nonlinear and complex, achieved through kernel tricks with linear, polynomial, radial basis, or sigmoid kernels.

\textbf{Naive Bayes (NB)}: NB is a probabilistic machine learning algorithm based on Bayes' theorem. It assumes feature independence within a dataset. The algorithm calculates the probability of a data point belonging to a certain class based on its feature values and the prior probability of that class. Subsequently, it assigns the class with the highest probability to the data point \cite{webb2010naive}.

\textbf{Decision Tree (DT)}: DT functions by recursively partitioning the data based on input features to minimize the impurity of resulting subsets. Each node in the tree signifies a decision based on a feature value, while the leaves represent the final prediction. Impurity is typically measured using metrics like Gini impurity or entropy \cite{myles2004introduction}. \textbf{J48} is a popular algorithm for building DTs.

\textbf{Random Forest (RF)}: RF operates by creating a set of DTs, where each tree is trained on a bootstrapped sample of data and a random subset of features. The final prediction results from aggregating the predictions of all trees, either by majority voting (classification problems) or averaging (regression problems) \cite{biau2016random}.

\textbf{Neural Network (NN)}: NN comprises interconnected neurons arranged in layers to process and transmit information. The input layer receives data, hidden layers perform computations, and the output layer produces the final prediction. NNs learn from large datasets, capturing complex nonlinear relationships between features \cite{bishop1994neural}. The most common type is the \textbf{\textit{feed-forward neural network}}, where information flows unidirectionally from input through hidden layers to the output. Examples include Multi-layer Perceptron (MLP), Convolutional Neural Network (CNN), and Radial Basis Function (RBF) networks.

\textbf{Deep Belief Network (DBN)}: DBN is an unsupervised machine learning algorithm consisting of multiple layers of restricted Boltzmann machines (RBMs). Each layer of RBMs is trained unsupervised to learn a compressed representation of input data \cite{hinton2009deep}. The output of one RBM layer serves as input to the next, and the entire network is fine-tuned using supervised learning, like backpropagation.

\textbf{RNN \& LSTM}: Recurrent Neural Network (RNN) processes sequential data by retaining a memory of past inputs \cite{grossberg2013recurrent}. It operates by processing each input in the sequence, updating the memory state, and producing an output fed back into the RNN for the subsequent time step. Long Short-Term Memory (LSTM) addresses the vanishing gradient problem by using a more complex memory cell that selectively retains or discards information from previous time steps \cite{yu2019review}.

\textbf{Variational Auto-encoder (VAE)}: VAE, a generative deep learning model for unsupervised learning \cite{diederik2014auto}, encodes input data into a lower-dimensional latent space and decodes this representation to generate new, similar data. VAEs use a probabilistic approach to training and can learn complex data distributions.

\textbf{Generative Adversarial Network (GAN)}: GAN comprises a generator network and a discriminator network \cite{creswell2018generative}. The generator learns to create new data resembling the training data, while the discriminator learns to differentiate between generated and real data. They are trained simultaneously in a min-max game where the generator attempts to deceive the discriminator while the discriminator aims to correctly identify generated data.


\textbf{Genetic Algorithms:} Genetic algorithms are a type of optimization algorithm that mimics the process of natural selection and evolution. They involve creating a population of potential solutions, evaluating their fitness based on a given objective function, and then selecting the fittest individuals to reproduce and create the next generation of solutions. This process is repeated iteratively, with each generation getting closer to an optimal solution  \cite{forrest1996genetic}.


\subsection{Taxonomy}\label{sec:taxonomyofadversarial}

\subsubsection{Definitions}Here, we define a list of terminology that will be used throughout the paper.\par 
\textbf{Adversary and Defender:} An adversary, also referred to as an attacker, aims to deceive the classifier into producing a misclassification. A defender, on the contrary, develops security defenses to protect against adversarial attacks.
\par
\textbf{Classifier:} A classifier is a machine learning model designed to make a classification decision based on input data. Examples of classifiers include Support Vector Machines (SVM) and K-nearest neighbors (KNN).
\par
\textbf{Adversarial Attack:} An attack by an adversary that aims to fool the classifier is called an adversarial attack. Adversarial attacks can occur at either the training or testing time. An evasion attack, which occurs at test time, involves modifying the test samples in such a way that they evade the detection of the classifier. Poisoning attacks, on the other hand, manipulate the training data with the intent to cause the misclassification of test data.

\begin{figure}[]
\centering
\includegraphics[width=0.58\textwidth]{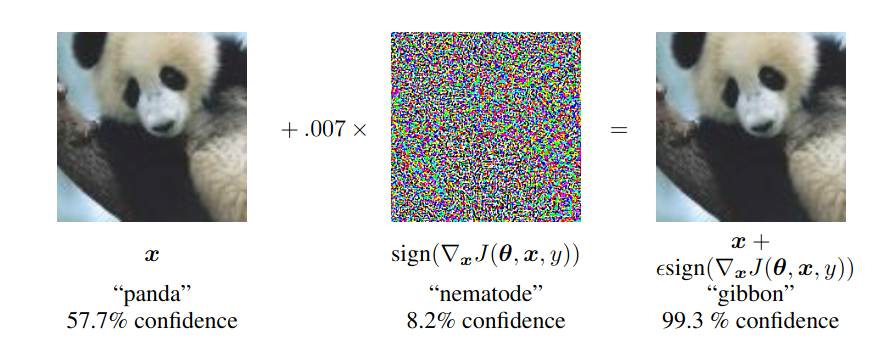}
\caption{Goodfellow et al. \cite{goodfellow2015explaining} demonstration of adversarial example generated using Fast Gradient Sign attack. The left image is correctly classified by GoogleNet \cite{szegedy2015going} as a panda. On adding an imperceptible noise, the image on the right, which looks similar to the original image, is misclassified as a gibbon.}
\label{fig:evasion_attack_goodfellow}
\end{figure}

\par 
\textbf{Adversarial Sample:} An adversarial sample is a sample modified by an adversary by adding a perturbation such that the model output is changed from the normal label. Figure \ref{fig:evasion_attack_goodfellow} demonstrates an example of an adversarial attack and adversarial sample. Adversarial example exhibits transferability property, which states that the adversarial example generated to fool a specific model can also fool other models \cite{szegedyintriguing, goodfellow2015explaining,papernot2016transferability}. 

\textbf{Perturbation:} A perturbation is a carefully crafted noise that is strategically added to the input data to fool the classifier. The manipulation can occur by adding or removing code from the app source code to misclassify the malware as benign. In the context of images, the perturbation vector can modify pixel values. The primary objective of an adversarial evasion attack is to identify the minimal perturbation that can modify the classifier decision on an input sample. This objective can be stated formally as shown in Eq. \ref{eqn:adver}:

\begin{equation}\label{eqn:adver}
\text{arg min}_{\delta x}||\delta x|| \quad \text{s.t.} \quad F(x + \delta x) = t
\end{equation}

\begin{figure}[]
\centering
\includegraphics[width=0.55\textwidth]{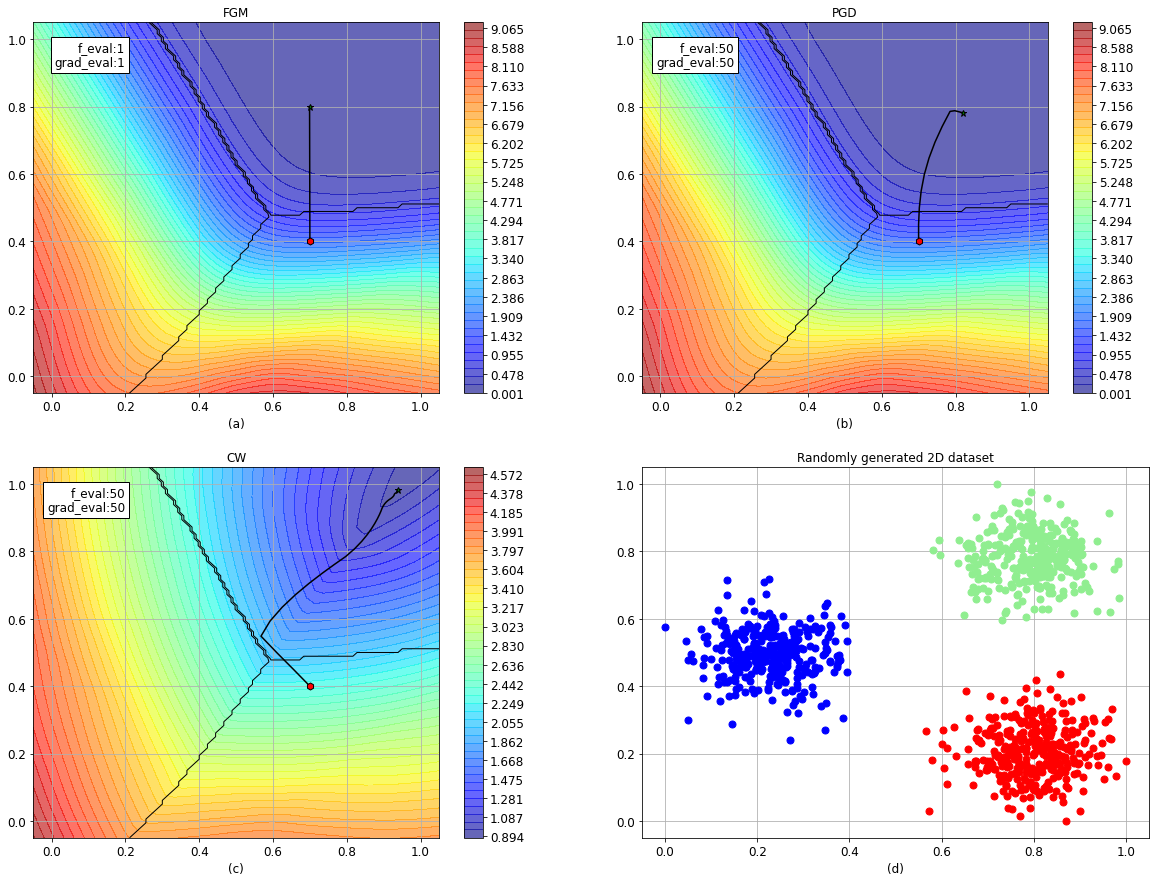}
\caption{Representation of evasion attack in a multi-class classifier using SecML \cite{melis2019secml} and Cleverhans \cite{papernot2018cleverhans}}
\label{fig:evasion_attack_cleverhans}
\end{figure}

Here, $ ||.|| $ represents the norm used to compare the original input and adversarial samples. Adversarial attacks commonly employ three major norms: $L_0$, $L_2$, and $L_\infty$. In $L_0$ norm, the distance between two samples is measured as the count of perturbed features where original sample differs from the modified sample. $L_2$ norm measures the Euclidean distance between two samples. In $L\infty$ norm, the similarity between two samples is measured by the maximum change to any of the input features. Various attack methods use different norms to generate perturbation vectors. Symbolically, $\delta x$ is the perturbation added to the sample $x$ to create an adversarial sample $x*=x+\delta x$, leading to misclassification as $t$. Heuristics, brute force, or optimization algorithms can be used to obtain such minimal perturbations. 

Figure \ref{fig:evasion_attack_cleverhans} shows how a test sample undergoes perturbation in feature space using three distinct attack types for inducing misclassification. The illustrative 2D dataset comprises 3 clusters of randomly generated points from a normal distribution. We train a 2-layer neural network for making multi-class classification and obtain an accuracy of $99\%$ on the test set. We then perform FGM, PGD, and CW attacks (see Section \ref{sec:AdversarialAttackImage} for details on the attacks) on the classifier to generate adversarial samples. Figures (a), (b), and (c) depict how an initial sample (represented by a red hexagon) belonging to one class is perturbed in its feature space and transformed to its adversarial sample (green star) for misclassification by the respective attacks. f\_eval and grad\_eval denote the number of function and gradient evaluations required to execute the attack.

\subsubsection{Threat Model}\label{threatModel}

Threats from an adversary are defined in terms of their objectives and capabilities, which form the core components and assumptions of threat modeling:
 
\textbf{1. Adversarial goal:} An adversary aims to fool the target model by producing misclassifications. In targeted misclassification, the adversary aims to produce a specific output label, whereas, in untargeted misclassification, the adversary is not concerned with the specific model output.

\textbf{2. Adversarial knowledge:} An adversary's knowledge about the target classifier can be partial (gray box), complete (white box), or zero (black box) on the training data, feature set, learning algorithm, parameters, and hyper-parameters.

\textbf{3. Adversarial capabilities} An adversary's capabilities define how they can fool the classifier. An evasion attack is performed at test time to change the classification decision. A poisoning attack is carried out by polluting the training data at train time. This survey paper focuses on white and black-box test-time adversarial attacks, called evasion attacks.

\subsection{Android Systems}\label{sec:androidfile}

\textbf{File Structure: }Android applications are packaged in APK file format for distribution and installation. The files in an APK package characterize the behavior and operation of an Android app (see Table \ref{APK_Table}). These files are used in classifier design and also modified by an adversary while performing adversarial attacks. We explain the files and directory below: 
\begin{enumerate}
    \item \textbf{Androidmanifest.xml:} This configuration file consists of a list of activities, services, and permissions used to describe the intentions and behaviors of an Android application. It is one of the majorly exploited files for evasion attacks. 
    \item \textbf{classes.dex:} Android applications are written in Java language. The Java source code is optimized and packed in a .dex file with references to any classes or methods used within an app.
    \item \textbf{resources.arc :} This is a pre-compiled file with information on application resources like design, strings, and images.
    \item \textbf{res:} This is a folder that contains application resources that are not compiled in resources.arsc file.
    \item \textbf{Meta-INF:} This directory is available in signed APKs with metadata information of Android apps like a signature. 
    
\end{enumerate}

\begin{table} []
\renewcommand{\arraystretch}{1}
\small
\caption{Structure of an Android Package Kit (APK)}
\label{APK_Table}
\centering
\resizebox{0.48\textwidth}{!}{%
\begin{tabular}{l c l}
\hline
File \& Directory & Type  & Description \\ \hline

META-INF & dir & APK metadata \\
assets & dir & Application assets\\
lib & dir & Compiled native libraries \\
classes.dex & file & App code in the Dex file format\\
res & dir & Resources not compiled into resources.arsc\\
resources.arsc & file & Precompiled resources like strings, colors, or styles\\
AndroidManifest.xml & file & Application metadata — e.g. Rights, Libraries, Services\\
\hline 
\end{tabular}}
\end{table}

\textbf{Malware Dataset}: The Drebin dataset  \cite{arp2014drebin}, which includes 5560 malware applications collected between August 2010 and October 2012, is widely used in Android malware research. Another popular dataset is AMD \cite{wei2017deep}, which contains 24650 malware applications collected between 2010 and 2016 and is organized into 135 families. The Canadian Institute for Cybersecurity offers various types of malware datasets\footnote{https://www.unb.ca/cic/datasets/index.html}. Androzoo \cite{allix2016androzoo}, which provides access to over 22 million Android APKs, is another notable resource in this field\footnote{https://androzoo.uni.lu/}.

\subsection{Malware Classifiers}\label{sec:malwareanalysis}

Malware encompasses any malicious software that exhibits malicious activity, e.g., trojan, worm, backdoor, botnet, spyware \cite{faruki2014android}. Trojans disguise as benign apps but perform malicious actions without the consent of the users (e.g., Fakeinst \cite{hidhaya2013detection}). Backdoor employs root exploits to allow other malicious apps to enter and perform activities (e.g., Kmin \cite{zhou2012dissecting}). Worm has the ability to duplicate itself and spread via network or removable devices (e.g., Obad \cite{backdoor}). Botnet creates a bot that allows remote control of the device (e.g., genome \cite{zhou2012dissecting}). Spyware monitors the functioning of other applications in the Android phone and sends the information to an adversary (e.g., Nickyspy \cite{zhou2012dissecting}). Ransomware locks the device from the user until a ransom amount is paid (e.g., FakeDefender.B \cite{fakedefend}). A malware classifier \textit{obtains} an Android application and performs malware analysis and malware detection to categorize it as benign or malware. 

\textbf{Malware Analysis:} Malware analysis is performed to understand the malware and extract its characteristic features \cite{payet2012static}. 
There are three types of analysis approaches: static, dynamic, and hybrid. \textit{Static analysis} does not require running the executable file and instead utilizes tools such as Androguard \cite{androguard}, APKTool \cite{apktool} or Dex2jar \cite{dex2jar} to decompile the APK file and extract features like permission (request and use), intent, component (activity, content provider), environment (feature, SDK, library), services, strings, API calls, data flow graphs, and control flow graphs. Extraction of static features is computationally inexpensive and fast; however, detectors based on static features have low accuracy when malware employs encryption or obfuscation technology \cite{karbab2018maldozer}. On the other hand, \textit{dynamic analysis} involves running the application in a virtual or real environment and monitoring its behavior to extract dynamic features. Common toolboxes for dynamic analysis are AndroPyTool \cite{martin2019android} and MobSF \cite{mobsf}.  Some dynamic features represent the intrinsic behavior that is resistant to app obfuscation \cite{wong2016intellidroid}. For example, system calls, app patterns, privileges, file access details, opening of services, information flows, and network traffic. \textit{Hybrid} methods leverage the advantages of both static and dynamic analysis by collecting features from both techniques.

\begin{figure}[h]
\centering
\includegraphics[width=0.78\textwidth]{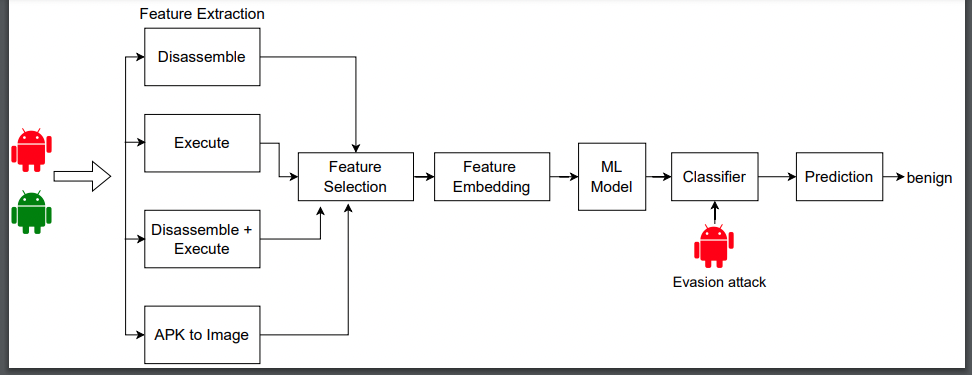}
\caption{A system diagram of an Android malware classifier with an evasion attack.}
\label{fig_malware}
\end{figure}

\textbf{Malware Detection:} Malware detection can either use signature-based methods or machine learning methods. Signature-based methods rely on a repository of known malware signatures, which are used to test whether a new application is malicious or not \cite{moser2007limits}. It is not an effective method since there is a need to keep updating the repository with signatures of malware. Machine learning-based models are trained on features extracted from Android apps using static, dynamic, or hybrid analysis (see Figure \ref{fig_malware}). An alternative approach to malware detection is to use image-based classifiers. In an \textit{image-based classifier}, static features from \emph{AndriodManifest.xml} and \emph{classes.dex} files are converted to RGB or gray-scale images while preserving the local and global features of the APKs \cite{nataraj2011malware}. These image features represent binaries of an Android app and are used to train a convolutional neural network (CNN) based malware classifier. Several ML-based classifiers are proposed in the literature to detect Android malware \cite{arp2014drebin, 
komogortsev2015attack, lindorfer2015marvin, saracino2016madam, su2016deep,hou2016deep4maldroid, yuan2016droiddetector, sun2016monet,wang2017detecting, kim2018multimodal, fan2018android,cai2018droidcat, onwuzurike2019mamadroid, arora2019permpair, han2020enhanced, li2018significant,han2020android}. These classifiers differ in their proposed approach of extracting features either using static, dynamic, or hybrid analysis. Below, we provide a review of the most popular malware classifiers in the literature:

One of the most well-known datasets is \textbf{Drebin} \cite{arp2014drebin}, which encompasses various static features from Android applications, including permissions, components, intents, and API calls. These features were employed in an SVM-based classifier to train a detection model. Wang et al. \cite{wang2014exploring} have also utilized permissions used by the application by ranking them in order of risk using three feature ranking methods: mutual information, correlation coefficient, and T-test. Later, they extract a set of risky permissions and use SVM, decision trees, and random forests to build a binary malware classifier. Another classifier, \textbf{Marvin} \cite{lindorfer2015marvin}, utilizes both static and dynamic features for training linear models. They extract static features, such as permissions and certificates, and dynamic features, like data leakages and network communication. Su et al. \cite{su2016deep} propose \textbf{DroidDeep}, which uses a deep belief network (DBN) to learn the malicious features and an SVM-based detector to classify the malware. \textbf{DroidDetector} \cite{yuan2016droiddetector} also employs a deep belief network for malware classification by conducting both static and dynamic analyses on the malware. Li et al. \cite{li2021robust} propose a malware detection approach that also provides defense for adversarial attacks using a combination of VAE and MLP.

In \textbf{Deep4MalDroid} \cite{hou2016deep4maldroid}, the authors introduce a unique dynamic analysis method called Component Traversal. This method generates system call graphs based on the code routines within Android applications. They then apply a deep learning framework to these graph features and conduct malware classification. Li et al. \cite{li2018significant} propose a framework called \textbf{SigPID} to identify malicious permissions from benign permissions. They identified 22 malicious permissions commonly utilized by malware and trained 67 different Machine Learning algorithms to evaluate the performance of the malware detector. Many researchers use permissions and API calls to group applications into benign and malware with SVM, random forest naive Bayes, J48, and neural network \cite{peiravian2013machine, qiao2016merging, alazab2020intelligent}. A multi-modal deep learning model was proposed by Kim et al. \cite{kim2018multimodal} for malware detection. The multi-modal approach was used to maximize the advantage of using multiple types of features: permissions, components, environment, strings, Dalvik opcode sequences, API call sequences, and opcode from Android APK files. Each type of feature vector was passed into a separate deep neural network. The final layers of each deep neural network were connected and fed as an input to the final deep neural network, which makes the classification. Chen et al.\cite{chen2018tinydroid} and Bai et al. \cite{bai2020famd} have also explored the use of opcode sequences as features for malware classification. 

\textbf{MaMaDroid} \cite{onwuzurike2019mamadroid} extracts the Control Flow Graph (CFG) of an Android application to use as features for classification. It extracts API calls from applications using static analysis and abstracts the calls to class, package, and family. The classification system analyzes the sequence of API calls and models the flow of information. Han et al. \cite{han2020enhanced} also utilize API calls to train SVM classifiers, but by representing them in a triple format comprising the API method name, argument, and return type to increase the dimensionality of the dataset. Vinod et al. \cite{vinod2019machine} and Xio et al. \cite{xiao2019android} extracted system calls as application features and trained random forest and LSTM models. Wang et al. \cite{wang2016trafficav} extracted network traffic of Android applications using TCP flow and HTTP packets and used those features to train a decision tree model. Mahindru and Singh \cite{mahindru2017dynamic} extracted permissions used by applications during dynamic analysis and used such features in the malware classifier. \textbf{DATDroid} is a comprehensive framework that monitors system calls, network packets, CPU usage, and memory usage during dynamic analysis. It employs a random forest model for malware detection based on these features\cite{thangavelooa2020datdroid}. 

\begin{figure}[!t]
\centering
\includegraphics[width=0.25\textwidth]{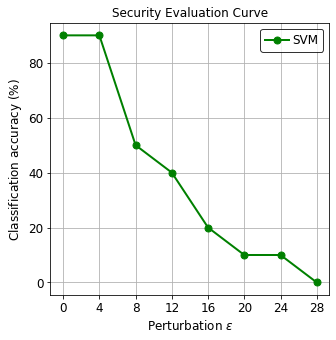}
\caption{A security evaluation curve showing the reduction in classification accuracy of an Android malware classifier as the perturbation $\epsilon$ (the number of feature modifications) is increased.}
\label{fig_securityEvaluation}
\end{figure}

\subsection{Example on Android evasion attack}\label{sec:exampleonandroid}
In this section, we describe an experiment to demonstrate an evasion attack on an Android malware classifier using SecML \cite{melis2019secml}. To construct the classifier, we use the Drebin dataset \cite{arp2014drebin}, comprising $12000$ benign and $550$ malicious samples. After splitting the dataset into training testing sets, we train a linear support vector machine (LSVM). This model achieves a classification accuracy of $95.04\%$ and an $89.09\%$ F1 score on the test set.
\par
To generate adversarial samples against the SVM classifier, we use a gradient-based attack \cite{biggio2013evasion}. Drebin represents the Android applications as one-hot encoded vectors of various permissions within the \textit{AndroidManifest.xml}. Therefore, during each iteration of the attack, we modify one feature of the Android application from $0$ to $1$. This adds a new feature to an Android application for its modification. Perturbation $\epsilon$ limits the number of features permitted for modification during the attack. To assess the robustness of the classifier against evasion attacks, we plot a security evaluation curve [See figure \ref{fig_securityEvaluation}] using SecML \cite{melis2019secml}. The curve plots the classification accuracy against the increasing perturbation $\epsilon$ value. Figure \ref{fig_securityEvaluation} shows how changing a mere $12$ number of features reduces the classification accuracy of the classifier by half. This simple example model demonstrates that an evasion attack on malware classifiers is a significant security threat.

\section{Understanding adversarial attacks}\label{sec:whyVulnerable}

Adversarial attacks exploit aberrations in the learned parameters of the classifier. Earlier, the hypothesis was that the presence of `blindspots' in neural networks caused adversarial vulnerability of classifiers \cite{szegedyintriguing}. Szedy et al. argued that the adversarial examples represented the low probability \emph{pockets} in the network manifold created due to insufficient training examples. If an adversarial sample existing in that manifold is supplied during test time, the model will fail to classify it. Such lower-probability pockets were the result of high non-linearity in deep neural networks.
\par
Contrary to the non-linearity hypothesis of \cite{szegedyintriguing}, Goodfellow et al. \cite{goodfellow2015explaining} have argued that the resemblance of neural networks to linear classifiers in high dimensional space is the reason behind a classifier's inability to determine adversarial samples accurately. If a linear model $y=\theta^{T}x$ receives a modified input $x'$ such that $x'=x+\delta$, the new activation $\theta^{T}x'$ in the classifier is expressed as:

\begin{equation}\label{eqn:linearAdver}
     \theta^{T}x'=\theta^{T}(x+\delta)=\theta^{T}x+\theta^{T}\delta
 \end{equation}
 
Equation \ref{eqn:linearAdver} shows that adding an adversarial perturbation of $\delta$ increases the activation of the model by $\theta^{T}\delta$. If the input is high-dimensional, a slight input perturbation can create a large perturbation in output.

However, Tanay et al.\cite{tanay2016boundary} argued that this linear view of neural networks poses several limitations primarily because the behavior of a neural network and a linear classifier is different. They verified that the activation function in the neural network units does not grow linearly, as explained in \cite{goodfellow2015explaining} by finding adversarial examples that resist adversarial perturbation in a linear classifier. Instead, they propose that adversarial examples exist when the class boundary lies close to data. Because of this, modification of points from the data manifold towards the boundary will cause the data to misclassify. Overfitting could be the rationale behind this behavior, and they argued that proper regularization could address this phenomenon.

\begin{figure}[!t]
\centering
\includegraphics[width=0.80\textwidth]{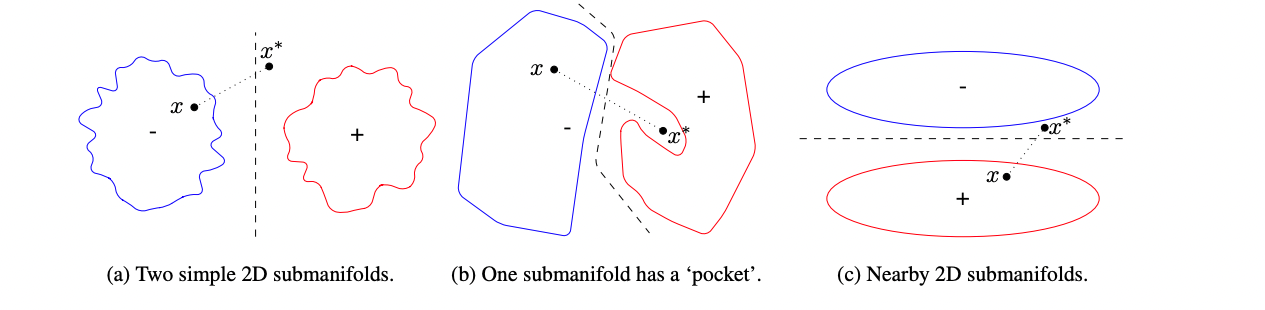}
\caption{Explaining the position of adversarial examples in data manifold following the assumption of Feinman et al. \cite{feinman2017detecting}.}
\label{fig_positionadversarial}
\end{figure}

Feinman et al. \cite{feinman2017detecting} based their detection of adversarial examples on the assumption that adversarial samples do not lie on the true data manifold and pose three different settings where adversarial samples could exist. As demonstrated in Figure \ref{fig_positionadversarial}, an adversarial sample $x^*$ is produced by moving a malware sample $x$ away from its data manifold (represented in blue) and across the decision boundary (dashed line). There are three possibilities for the position of such an adversarial sample: (a) $x^*$ is far from the data manifold of benign samples (represented in red). (b) the input space has low probability pockets, and adversarial sample $x^*$ falls into such pockets. (c) the adversarial sample $x^*$ lies near the decision boundary close to both the data manifolds. Recent work by Ilyas et al. \cite{ilyas2019adversarial} claimed that adversarial vulnerability is the result of the model's sensitivity toward feature generalization. They argued that classifiers train to maximize accuracy regardless of the features they use. Because of this, the model tends to rely on  `non-robust' features for making decisions allowing possibilities of adversarial perturbations. This hypothesis also explains the presence of adversarial transferability.

\par 
Though the verdict is yet not unanimous in explaining the cause behind adversarial vulnerability, this new hypothesis on adversarial vulnerability being the intrinsic property of data and the result of our goal for generalization has paved the way for new research in building defenses \cite{zhang2019defense, kim2020adversarial}.
\section{Evasion attacks and defenses: Foundation}\label{sec:AdversarialAttackImage}
We begin by explaining foundational work in adversarial machine learning commonly demonstrated for computer vision classifiers. We discuss adversarial attacks in Android malware in Section \ref{sec:AdversarialMalware}. 

The following symbols are used in the subsequent sections:
\begin{itemize}
    \item $X$: Set of d-dimensional input, $x \in X$ is a sample 
    \item $Y$: Set of label of input dataset, $y \in Y$ is a sample 
    \item $F$: A trained classifier (model)
    \item $\theta$: Model parameters 
    \item $J(\theta, X, Y)$: Loss function of the classifier 
    \item $\epsilon$: Perturbation parameter
    \item $\nabla_x$: Gradient with respect of $x$
\end{itemize}

\subsection{Evasion attacks}
\textbf{L-BFGS \cite{szegedyintriguing}:} In 2014, Szegedy et al. formulated the evasion attack as an optimization algorithm and proposed a solution using a box-constrained limited memory Broyden–Fletcher–Goldfarb–Shanno algorithm (L-BFGS) \cite{liu1989n0cedai}.

\begin{equation}\label{eqn:lbfgs}
    \textnormal{min}_\delta \Big ( c||\delta||_2+\textnormal{loss}_{F}(x+\delta,t) \Big )
\end{equation}

subject to $x+\delta \in [0,1]^n$; $t$ is the target label. The goal of the optimization in Equation \ref{eqn:lbfgs} is twofold. The first is to perform a linear search to find the constant $c$ that locates a minimally perturbed sample. The second is that the modified image $x+\delta$ must still normalize between $0$ and $1$. This method can be computationally slow and not scale to large datasets. Additionally, this paper presents two compelling properties of adversarial examples. First, the adversarial examples generated from a classifier can often evade other classifiers trained with the same training dataset, even with different model architectures and parameters. Second, they could even fool models trained on entirely different, disjoint training datasets. These two properties of adversarial samples came to be known as ``cross-model generalization" and  ``cross-training set generalization" of adversarial attacks.
\par 
\textbf{FGSM \cite{goodfellow2015explaining}:} Unlike Szegedy et al. \cite{szegedyintriguing} that optimizes the perturbation for $L_2$ distance and solves an optimization problem, Goodfellow et al. optimize perturbation for  $L_\infty$ metric and assumes a linear approximation of the model loss function. The optimal max-norm perturbation uses the following expression: 
 \begin{equation}\label{eqn:fgsm}
     \delta=\epsilon * \textnormal{sign}(\nabla_{x}J(\theta,x,y))
 \end{equation}
 Equation \ref{eqn:fgsm}, also called the fast-gradient sign method (FGSM) for an un-targeted attack, finds the perturbation by increasing the local linear approximation of the loss function. FGSM is  `fast' because the loss gradient computes once and modifies the input perturbation of single-step size $\delta$. 
 
\par

\textbf{Iterative FGSM (BIM) \cite{kurakin2016adversarial}:} Kurakin et al. introduced a method that iteratively applied the Fast Gradient Sign Method (FGSM). It involves calculating the sign of the gradient of the loss function with respect to the input, multiplying it by the step $\alpha$, and clipping the output to $\epsilon$ so that the transformed input is on the $\epsilon$ neighborhood. This process is repeated with multiple smaller steps of $\alpha$. The number of iterations is heuristically selected to be between $\textnormal{min}(\epsilon + 4, 1.25~\epsilon)$.
 \begin{equation} \label{eqn:iterativegradient}
     x*=x_{n-1}-\textnormal{clip}_{\epsilon}(\alpha~\textnormal{sign}(\nabla xJ(\theta,x_{n-1},y))
 \end{equation}
 \par 
\textbf{Projected Gradient Descent (PGD) \cite{madry2018towards}:} PGD is another iterative method similar to BIM. However, unlike BIM, to find the local maximum loss value of the model, the PGD attack starts from a random perturbation in $L_p$-ball around the input sample.
 \begin{equation} \label{eqn:iterativegradient1}
  {x}^*={x}_{n-1}-\textnormal{clip}_{\epsilon}(\alpha~\textnormal{sign}(\nabla xJ(\theta,{x}_{n-1},y))
 \end{equation}

\textbf{Jacobian-based Saliency Map (JSMA) \cite{papernot2016limitations}:} Papernot et al. generate a novel adversarial attack using $L_0$ distance metric to minimize perturbation of the input. They utilize salient maps to select a subset of input features to be modified to produce misclassification. A saliency map of an image quantifies how each input feature influences the model's prediction. 
\par 
\textbf{C\&W \cite{carlini2017towards}:} Carlini et al. devise one of the strongest attacks, using $L_0$, $L_2$ and $L_{\infty}$ distance metrics. The C\&W attack is similar to the attack proposed by Szegedy et al. \cite{szegedyintriguing}; however, the difference lies in the use of the objective function. While L-BFGS used cross-entropy loss, the C\&W attack uses margin loss. Here, we formally define this adversarial attack:
\begin{equation}
    \textnormal{minimize}~||\delta||_{p}+\lambda'.F(x+\delta)
\end{equation}

such that $x+\delta\in [0,1]^n$. This optimization is reformulated with the following function:

\begin{equation}\label{eqn:cwfx}
    F(x*)=[\textnormal{max}_{i\neq t} Z(x*)_{i}-Z(x*)_{t},-k]^{+}
\end{equation}

Here, $k$ encourages the optimizer to find an adversarial example with high confidence, and $Z(.)$ are logits. 

\textbf{DeepFool \cite{moosavi2016deepfool}:} DeepFool is an iterative $L_2$-attack designed to produce minimally perturbed samples. For a linear binary classifier, the minimal perturbation is obtained by finding the projection of a sample $x$ on the classifier's decision boundary, a hyperplane $F:\theta^{T}x+b=0$. This projection is given by $-\frac{F(x0)}{||\theta||_{2}^{2}}\theta$. For a general curved decision surface, a linear approximation is obtained using a small linear step of $\nabla F(x)$ at each iteration $x_i$. Using the closed form solution of perturbation $\delta$ for linear case, the perturbation $\delta$ for general case is expressed as: 
\begin{equation}\label{eqn:deepfool}
    \delta_{i}=-\frac{F(x_i)}{||\nabla F(x_i)||_{2}^{2}}\nabla F(x_i)
\end{equation}
The modified sample $(x*)$ is obtained by adding $\delta$ from equation \ref{eqn:deepfool} to $x_i$.

\textbf{Practical black-box attack \cite{papernot2017practical}:} Papernot et al.  demonstrate the first black-box attack that fooled a remotely hosted model (Oracle). By repeating three steps: querying Oracle for a label, training a substitute model, and generating a synthetic dataset, an adversary can create a substitute model that can mimic the Oracle decision behavior. Adversarial samples are generated using any attacks on the substitute model. Such samples can evade the classifier of the Oracle following the transferability property of adversarial attacks.

\par

\textbf{Universal Adversarial Perturbation (UAP) \cite{moosavi2017universal}:} These perturbations are input-agnostic, meaning that instead of designing perturbations for each individual input image, this method discovers a single perturbation that can cause misclassification in the majority of images within a given dataset. UAP aggregates perturbations for several input instances and conducts a sample-specific attack to construct a sample-agnostic UAP. 
\par 
\textbf{Generative attack \cite{zhao2018generating}:} Zhao et al. use GAN \cite{goodfellow2014generative} for producing adversarial samples. The generator is trained to learn a data-generating distribution mimicking input $x$ from random vectors $z \in R^d$. Using an inverter network ($I$), the input sample is mapped to the generator labeled with incorrect output. Mathematically, the adversarial sample is given by:
\begin{equation}
    x*=G_\theta (z*) ~ \textnormal{where} ~ z*=\textnormal{argmin}_{z'}||z'-I_\gamma (x)||
\end{equation}
such that $F(G_\theta (z'))\neq f(x)$. This approach utilizes input instances for creating adversarial samples and does not depend on an attack algorithm.
\par 
\textbf{Real-world adversarial attacks:} Few works have demonstrated the vulnerability of deep learning models in real-world environments. Sharif et al. \cite{sharif2016accessorize} designed adversarial eyeglasses using an optimization framework similar to L-BFGS that can evade face recognition systems. Evtimov et al. \cite{evtimov2017robust} proposed Robust Physical Perturbation ($RP_2$), which demonstrated that adding posters and stickers to existing real-life images like traffic signs can easily fool a deep learning model. Similar to this, Brown et al. \cite{brown2017adversarial} came up with an input-agnostic adversarial patch that causes a classifier to misclassify an input when the patch is present in the image. 

\subsection{Adversarial defenses}
Defending against an adversarial attack either involves detecting an adversarial sample and rejecting it, called adversarial detection, or building a robust classifier that is, by design, resilient against evasion attacks. Several attack algorithms have shown that no defense methods are robust enough to handle all kinds of attacks \cite{carlini2017towards}. An adversary can always exploit the limitations of defenses by carefully crafting new attacks. Here, we explain some popular defense methodologies proposed in the literature.
\subsubsection{Improving model robustness}\label{sec:improvingmodelrobustness}
There are various approaches that aim to make a deep learning model more robust against adversarial attacks. We discuss some major techniques below: 
\par 
\textbf{Adversarial training:} Adversarial training \cite{goodfellow2015explaining} is a popular defense method using a mixture of normal and adversarial examples for training. The additional set of adversarial examples provides information regarding adversarial  `outliers' to improve model performance against adversarial attacks. Equation \ref{eqn:adversarialtraining} shows the objective function for adversarial training where $J'$ is the modified loss function.
\begin{equation}\label{eqn:adversarialtraining}
    J'=\lambda' J(\theta,x,y) + (1-\lambda')J(\theta,x+\epsilon ~ \textnormal{sign}(\nabla_{x}~J(\theta,x,y))
\end{equation}
\par 

Adversarial training can overfit to perturbation samples used for training \cite{akhtar2018threat} 
and is unsuitable for scalability \cite{kurakin2016adversarial}. An ensemble adversarial training method can address the first limitation \cite{tramer2017ensemble} where the training set is augmented with different perturbations obtained from attacks on several pre-trained models. Kurakin et al. address the second limitation in \cite{kurakin2016adversarial} by proposing an improved version of adversarial learning suitable for large-scale datasets like ImageNet. The input datasets are grouped into batches of original and adversarial samples before training. 
\par 
\textbf{Input regularization:} Ross et al. \cite{ross2018improving} formulate that networks equipped with input gradient regularization are more robust to adversarial attacks than unregularized networks. Using this formulation, they propose the following objective function:
\begin{equation}
    \theta*=\textnormal{argmin}_\theta J(y,y')+\lambda'||\nabla_xJ(y,y')||_2^2
\end{equation}
This regularization ensures that the output prediction does not change drastically on a slight input modification.

\textbf{Defensive Distillation:} Hinton et al. \cite{hinton2015distilling} proposed distillation for deep learning to utilize the knowledge acquired by a large and complex neural model and transfer it to a smaller model. Papernot et al. \cite{papernot2016distillation} apply the intuition of distillation for improving the resilience of deep neural networks against adversarial samples. In distillation, soft labels train a classifier instead of high-confidence, hard labels. It reduces the sensitivity of the model with respect to variations in input (Jacobian) which are exploited for generating adversarial samples in \cite{szegedyintriguing, goodfellow2015explaining, papernot2016limitations}. 
\par 
\textbf{Stochastic Activation Pruning}: In their work, Dhillon et al. \cite{dhillon2018stochastic} conduct random pruning of some model activations during the forward pass. Activations with larger magnitudes are preserved, and after the pruning, the remaining activations are scaled up to normalize the input's dynamic range to the subsequent layers. The activation pruning introduces stochasticity, making it challenging for an adversary to efficiently compute gradients and generate potent adversarial samples.
\par 
\textbf{Denoising block}: Adversarial attacks often involve adding small amounts of noise to the input features to deceive the network into making incorrect predictions. Feature denoising can help to make the network more robust to these attacks by removing the noise from the input features before they are processed by the network. This technique involves adding a noise filter to the network that removes noisy input features before they are processed by the network. Xie et al. \cite{xie2019feature} proposes filters to denoise the feature map in a network. They design filters like mean filtering, median filtering, non-local means and bilateral filtering and add them as a denoising operation at any level of the network. The technique of feature denoising can enhance a network's resilience to attacks by eliminating noise from input features before they are processed. This approach typically involves the addition of a noise filter to the network during training \cite{xie2019feature}.

\subsubsection{Adversarial detection} Instead of making a model robust, adversarial detection methods use properties of benign and adversarial images to detect adversary samples. 
\par 
\textbf{KD+BU \cite{feinman2017detecting}}: Feinman et al. proposed two metrics to perform adversarial detection: Kernel Density (KD) and Bayesian Uncertainty (BU)
Estimates. KD estimates measure whether a given input sample is far from a class manifold, while BU detects input samples that are near the low-confidence region. They learn measures for benign and adversarial samples and fit a logistic regression model that acts as the detector.
\par 
\textbf{Local Intrinsic Dimensionality (LID) \cite{ma2018characterizing}:} Ma et al. propose the use of LID to estimate the space-filling capability of the region surrounding a given input sample. This is accomplished by calculating the distance distribution of the sample and its neighboring samples across multiple layers. The resulting LID scores are then used to train a logistic regression classifier capable of detecting both benign and adversarial samples.

\par 
\textbf{Feature Squeeze \cite{xufeature}:} Xu et al. apply two feature squeezing methods to test images: reduction of color depth and spatial smoothing of images. An image is passed through a classifier during testing to generate an output probability vector. The same test image is passed through feature squeezing networks and then to the same classifier, producing a probability vector. The difference between those two prediction vectors is compared with thresholds to detect the adversarial sample. 
\par
\textbf{Maximum Mean Discrepancy (MMD) \cite{gretton2012kernel}}: MMD is a distance metric that computes the maximum mean discrepancy between clean and adversarial examples to confirm whether a given test sample was drawn from a normal training data distribution or not.
\par 
\textbf{MagNet \cite{meng2017magnet}}: Magnet uses a detector network to discard such samples located far away from the data manifold learned by the classifier using training samples. If the adversarial sample is close to the boundary, then the reformer network transforms the adversarial sample into an original sample. They use auto-encoder to achieve this. 
\par 
\textbf{Deep Neural Rejection (DNR) \cite{sotgiu2020deep}}: DNR uses the output of the N-last layers of a deep learning model to train N-SVM classifiers.  The output of these N-SVM classifiers is again fed into another SVM classifier for detecting adversarial samples if the maximum confidence probability is less than a learned threshold.
\par 
\textbf{Network invariant approach \cite{ma2019nic}:}  Ma et al. monitor the provenance (explains the instability of activated neurons) and activation channel (explains the change in activation values) in a neural network for benign and adversarial samples by building a set of models for individual layers. The detector is jointly trained with the output of all the intermediate models. A given input is rejected as adversarial if it is classified as out-of-distribution from the provenance and activation channel. 
\par 
\textbf{Feature attribution-based detection:} Explanation methods can explain why a DNN made a particular decision and identify the key features in the input that led to that decision  \cite{ribeiro2016should}. Recent research has explored the application of feature-attribution-based explanation methods in detecting adversarial examples \cite{yang2020ml,wang2020interpretability}. For a given input instance, a feature attribution method gives a vector of scores with the same shape as the input. Feature attribution-based detectors extract such feature attribution for benign and adversarial samples and train supervised classifiers.

\section{Evasion attack and defenses: Patterns in android malware classifiers}\label{sec:AdversarialMalware}
In the context of malware, an evasion attack involves an adversary meticulously modifying a malicious application in such a way that the application evades the detection of a malware detector. An input like a malware application can be represented in two forms: problem-space objects, where the object lies in original input space corresponding to real-world applications, and feature-space objects, where the input objects have undergone transformation into a format suitable to process by the classifier. Let us consider $Z$ to be a problem space object (e.g. android APK file), $z \in Z$ be any object with a label $y \in Y$, $\phi: Z \rightarrow X \subseteq R^n$ be the feature mapping function that transforms the problem space object $z \in Z$ to an n-dimensional feature vector $x \in X$, such that $\phi(z) = x$. A machine learning classifier $f(.)$ produces a prediction $f(x)=y*$. This feature map function is not invertible in software. There is no straightforward transformation from feature-space object to problem-space object. Hence, an evasion attack introduces additional constraints in malware such that after performing an attack in feature space, we can transform the feature object into a real application.

\textit{\textbf{Evolution in Adversarial Machine Learning}}- Adversarial machine learning has significantly evolved from the initial advances in image datasets to more complex domains such as software encompassing binary-formatted Android Package Kit (APK) files, source code, and API calls. However, with this evolution in the dataset, adversarial machine learning introduces unique challenges in designing effective attacks and defenses \cite{pierazzi2020intriguing}. The major challenges are highlighted below:

\begin{enumerate}
\item Feature representation is fixed in image classifiers since it is constrained to pixels and, therefore, offers a limited attack surface. Whereas malware lacks a standardized feature representation, enabling adversaries to have greater flexibility in manipulating applications to circumvent detection by a classifier. This diversity of feature representations poses a significant challenge in designing robust defenses.

\item A perturbed image can be readily assessed through visual inspection. However, it is challenging to evaluate the impact of perturbations, and if it has altered the functionality and intrinsic property of an application.

\item The adversarial sample of a malware application should be able to execute successfully. The perturbations applied to the sample should not disrupt the underlying code structure or break the code structure of the application since this would render the application inoperable and potentially expose the attack. Maintaining functionality while deceiving the classifier is a key challenge in malware evasion attacks.
\end{enumerate}

\subsection{Evasion Attacks}

\begin{table*}[]
\centering
\caption{Adversarial attacks on android malware classifier covered in this paper. Acronyms: LR (Logistic Regression), DT (Decision Tree), RF (Random Forest), SVM (Support Vector Machine), MLP (Multi-layer Perceptron), DNN (Deep Neural Network), GBM (Gradient Boosting). G: (Gray box attack), B (Black box attack), W (White box attack)}
\label{tab:advAttacks}
\resizebox{\textwidth}{!}{%
\begin{tabular}{llclll}
\hline
\multicolumn{1}{c}{\textbf{Attacks}} & \multicolumn{1}{c}{\textbf{Year}} & \textbf{Transparency} & \multicolumn{1}{c}{\textbf{Dataset}} & \multicolumn{1}{c}{\textbf{Target model}} & \multicolumn{1}{c}{\textbf{High level summary}} \\ \hline
Grosse et al. \cite{grosse2017adversarial} & 2017 & W & Drebin & DNN & Compute Jacobian matrix of model and perturb feature in AndroidManifest.xml \\
Hu et al.  \cite{hugenerating} & 2017  & B & Web\footnote{https://malwr.com/} & RF, LR, DT, SVM, MLP & Employ the generative capability of a generative adversarial network (GAN) \\
Yang et al. \cite{yang2017malware} &  2017 & B & Drebin, VirusShare, Genome & KNN, DT, SVM, RF & Uses a program transplantation method to adds byte-codes to malware \\
Rosenberg et al. \cite{rosenberg2018generic}  & 2018 & B & VirusTotal & DNN, RNN & Perform mimicry attacks to mimic the system calls  of the benign code. \\
Shahpasand et al. \cite{shahpasand2019adversarial}  & 2019 & B & Drebin & SVM, DNN, RF, LR & Employs GAN architecture with threshold on  generated distortion \\
Chen et al. \cite{chen2019android} & 2019 & G & Drebin, AndroZoo & SVM & Adds perturbations to both  AndroidManifest.xml\} and  classes.dex \\
Liu et al. \cite{liu2019adversarial} & 2019 & B & Drebin & DNN, LR, DT, RF & Compute perturbation using genetic algorithm \\
Pierazzi et al.\cite{pierazzi2020intriguing}  & 2020 & W & Drebin & SVM, SecSVM & Uses software transplantation to extract benign slices to insert into malware \\
Li et al.  \cite{li2020adversarial} & 2020 & W & Drebin, Androzoo & DNN & Propose an ensemble attack on deep neural networks  for malware classification \\
Cara et al. \cite{cara2020feasibility} & 2020 & W & Drebin & MLP &  Studies whether adversarial samples can be created by injection of API calls. \\
Berger et al. \cite{berger2021crystal}  & 2021 & G, B & Drebin & SVM, Sec-SVM, DNN & Proposes three different attacks that modify manifest file in android APKs \\
Wang et al. \cite{wang2021exposing} & 2021 & W & Drebin & GBM, SVM, RF, DNN & Proposes an explainability-guided framework for generating adversarial attacks \\
Labaca et al. \cite{labaca2021realizable} & 2021 & W & Drebin & LR, DNN & Demonstrate that universal perturbations are possible  for malware classifiers \\
Zhao et al. \cite{zhao2021structural} & 2021 & W & Androzoo, Drebin & SVM, DNN & Designs an optimization model integrated with reinforcement learning (HRAT) \\
Bostani et al. \cite{bostani2021evadedroid} & 2022 & B & AndroZoo & SVM, Sec-SVM & Uses software transplantation technique to insert codes into malware \\
Berger et al.\cite{berger2022mamadroid2} & 2022 & G,W & AndroZoo & SVM, RF & Design attacks that manipulate the CFG of the MaMaDroid classifier \\ \hline
\end{tabular}%
}
\end{table*}

\subsubsection{Feature-space Attacks} In feature space attacks, the adversary modifies the feature space object $x$ into $x*$ so that the model prediction is evaded. However, these attacks come with constraints that ensure that the transformed feature-space object still functions as intended \cite{aryal2021survey}. 

Grosse et al. \cite{grosse2017adversarial} employed a feature selection method for producing adversarial samples of Android malware targeting a feed-forward neural network. They compute the gradient of the classifier $F$ with respect to input $x$ and then select a perturbation $\delta$ that maximizes the gradient with respect to $x$, i.e., pick the feature $i$ such that $i=arg max_{j\in[1,m],x_{j}=0} F_0 (x_j)$. This approach was proposed by Papernot et al., with JSMA \cite{papernot2016limitations}. In order to address the constraints for perturbation in malware, the authors only add features in \emph{AndroidManifest.xml} file and restrict the number of features added.

Li et al. \cite{li2020adversarial} propose an ensemble attack on neural networks for the modification of a malware sample with multiple attacks and manipulation sets. The attack set comprises of PGD \cite{kurakin2016adversarial}, Grosee \cite{grosse2017adversarial}, FGSM \cite{goodfellow2015explaining}, JSMA \cite{chen2019android}, salt and pepper noise \cite{schott2018towards}, and five obfuscation attacks. The manipulation sets put constraints on the permitted perturbation of the features of Android apps.

Hu et al. \cite{hugenerating} employ GAN to produce adversarial malware samples to attack a black-box classifier. This method trains the generator to mimic the distribution of malware samples and produce new adversarial samples. The target classifier labels both adversarial and benign samples, which are used as a training dataset for a substitute model for mimicking the target black-box classifier. The generator of GAN modifies the generated malware samples based on the prediction from the substitute model. To address the constraints on malware, they only add features to the \emph{AndroidManifest.xml}. Shahpasand et al. \cite{shahpasand2019adversarial} also implemented GAN to generate adversarial samples. The generator network of GAN learns benign sample distribution while producing perturbations to evade the classifier. The discriminator network enhances the perturbation by ensuring that adversarial counterparts are similar to benign samples.

\subsubsection{Problem-space Attacks} In problem space attacks, the adversary finds a sequence of transformations that mutates the problem space object $z$ to $z*$ so that the model prediction is evaded. Problem-space attacks also must satisfy constraints like available transformations, preserved semantics, robustness to preprocessing, and plausibility \cite{pierazzi2020intriguing}. 

Problem space attacks can be further classified into three forms of attacks: camouflage attacks use obfuscation or encryption to conceal the app information \cite{demontis2017yes}. The second type of attack adds noise to the applications \cite{pierazzi2020intriguing, cara2020feasibility}. Noise in malware refers to additional code or stub injection (uncalled functions). The third approach manipulates the app flow by performing app flow analysis using function outlining or inlining. An adversary replaces a function call with the entire function body in function inlining. In function outlining,  an adversary breaks a function into many smaller functions \cite{chen2019android}.

Chen et al. \cite{chen2019android} propose android HIV for adding perturbations to  \emph{AndroidManifest.xml} and  \emph{classes.dex} files. In order to attack the MaMaDroid classifier \cite{onwuzurike2019mamadroid}, they extract features from the \emph{classes.dex} file as Control Flow Graph (CFG). These features are represented as transition probabilities between different API calls and are used in building a Markov chain. To generate an attack, they insert API calls in the original smali code to modify this Markov chain. They modify the C\&W attack \cite{carlini2017towards} and JSMA attack \cite{papernot2016limitations} for this purpose. 

Liu et al. \cite{liu2019adversarial} propose a framework based on genetic algorithms to perform black-box adversarial attacks. Android applications are perturbed by adding specific permissions to AndroidManifest.xml, similar to Grosse et al. \cite{grosse2017adversarial}. The fitness function of the genetic algorithm searches for the optimum number of features to add that leads to an adversarial sample. 

Similar to surrogate model training by Papernot et al. \cite{papernot2017practical}, Rosenberg et al. \cite{rosenberg2018generic} propose a GADGET framework that generated adversarial examples without access to malware source code in black-box settings. Authors perform mimicry attacks to mimic the system calls of the benign code. Then, they train a surrogate model that behaves the same way as the target model by querying the black-box model with synthetic input values. Bostani et al. \cite{bostani2021evadedroid} use automated software transplantation to prepare actions by evaluating closely related benign apps of malware. Then, the code slices are added from the benign files to the malware app so that the classifier can be fooled into making the benign classification. Yang et al. \cite{yang2017malware} propose Malware Recomposition Variation (MRV) that systematically performs the semantic analysis of an Android malware to compute suitable mutation strategies for the application. Using such strategies, a program transplantation method \cite{barr2015automated} adds byte-codes to existing malware to create a new variant that successfully preserves the maliciousness of the original malware. Since they perform a semantic analysis of the malware to identify feature patterns before feature mutations, the adversarial sample is less likely to crash or break and preserve the maliciousness.
\par 
Pierazzi et al. \cite{pierazzi2020intriguing} investigate the problem-space attacks and propose a set of formalization by defining constraints on the adversarial examples. They use this formalization to create adversarial malware in Android on a large scale. They define the following constraints so that any transformation $T$ on problem-space object  $z \in Z$ leads to a valid and realistic object: the transformation must preserve the application semantics, the transformed object must be plausible, and the transformed object must be robust to pre-processing of a machine-learning pipeline. Perturbation in the problem-space attack is obtained using a problem-driven or gradient approach. In a problem-driven approach, the optimal transformation is performed empirically by applying random mutation on object $z$ using Genetic Programming \cite{xu2016automatically} or variants of Monte Carlo tree search \cite{quiring2019misleading} and learning the best perturbation for misclassification. The gradient-driven method approximates a differentiable feature mapping similar to MRV \cite{yang2017malware}. Cara et al. \cite{cara2020feasibility} study whether adversarial samples can be created by the injection of system API calls. They perform problem-space attacks by defining constraints similar to Pierazzi et al. \cite{pierazzi2020intriguing}. 
\par 
Berger et al. \cite{berger2021crystal} present three different evasion attacks that modify the manifest files in Android APKs: MB1, MB2, and MB3. MB1 finds the requested permission section to switch to adversarial permission tags. MB2 attack observes the permission family statistics in parallel to the manifest file and changes the manifest file accordingly by picking a permission family. MB3 attack blindly changes all of the uses-permission tags to adversarial tags. In another evasion attack, Berger et al. \cite{berger2020evasion} conceal the malware appearance of malicious APKs. The attacker searches for strings in the smali code that are included in the malicious report and replaces the string with its base64 encoding. They also evaluate whether the app’s functionality or malicious content has changed and claim that none of the prior works ensure that maliciousness is preserved.

Wang et al. \cite{wang2021exposing} propose an explainability-guided framework for generating adversarial attacks. To do so, they introduce the concept of Accrued Malicious Magnitude (AMM) to identify malware features that should be manipulated to maximize the likelihood of detection evasion using SHAP values \cite{lundberg2017unified}. Zhao et al. \cite{zhao2021structural} design a Heuristic optimization model integrated with a Reinforcement learning framework (HRAT). HRAT includes four types of graph modifications, viz. inserting nodes, deleting nodes, adding edges, and reconnecting nodes. These modifications correspond to manipulation on Android applications. However, these attacks are only suitable for function call graph (FCG) based classifiers. Bostani et a. \cite{bostani2021evadedroid} present a practical evasion attack, EvadeDroid, that attacks Android malware detectors in black-box settings. EvadeDroid first prepares a collection of functionality-preserving transformations to be applied to malware so that they are morphed to act as benign. The manipulation occurs via a query-efficient optimization algorithm. 

Labaca et al.\cite{labaca2021realizable} explore the challenges and strengths of universal adversarial perturbations (UAP) in malware and demonstrate that UAP is possible for malware classifiers by designing sequences of problem-space transformations that induce UAPs in the corresponding feature space embedding. They evaluate their attacks on the DREBIN classifier. In MaMaDroid 2.0 \cite{berger2022mamadroid2}, the authors propose attacks that manipulate the CFG of the MaMaDroid classifier with Structure Break Attack. Given malicious APKs, this attack manipulates the structure of each APK by changing the smali code files and manifest files. Only those elements that are part of the MaMaDroid feature set are changed.

\subsection{Adversarial Defenses}
\subsubsection{Improving model robustness}

Demontis et al.\cite{demontis2017yes} propose a classifier with uniform feature weight. This design approach forces the attacker to modify multiple features to change a classifier decision. They call this approach an adversary-aware design approach. They modify the Drebin \cite{arp2014drebin} classifier by introducing a regularization on feature weights and imposing an upper and lower bound constraint. This secure SVM classifier is also known as Sec-SVM in literature, given by Eqn. \ref{eqn:secsvm}.

\begin{equation}\label{eqn:secsvm}
    \textnormal{min}_{w,b}L(D,f)=\frac{w^{T}w}{2}+C\sum_{i=1}^{n}\textnormal{max}(0,1-y_{i}f(x_{i}))
\end{equation}
where, $w^{lb}_k \leq w_k \leq w^{ub}_k$. 
\par 

\par  
Grosse et al.\cite{grosse2017adversarial} investigate the performance of defense methods introduced in the context of computer vision in malware detection. They experiment with defensive distillation \cite{papernot2016distillation} and adversarial training \cite{goodfellow2015explaining}. 
Overall, adversarial training improved the robustness of a malware classifier against adversarial attacks. However, the number of adversarial examples used in the training had a significant impact on the robustness, and the model was robust only against such attacks whose samples were used in adversarial training. 
Defensive distillation had insignificant improvement in the defense. 
\par 
Yang et al. \cite{yang2017malware} also evaluate different defense mechanisms to improve the robustness of classifiers. Using adversarial training, they were able to improve the robustness of classifiers. They also evaluate weight bounding, similar to Demontis et al. \cite{demontis2017yes}, to improve the robustness of the malware classifiers. However, this defense mechanism performed worse than adversarial training. Chen et al. \cite{chen2019android} evaluate adversarial training and ensemble learning methods for adversarial defense. Adversarial training was able to improve model robustness, but it relied on the prior availability of adversarial samples. In ensemble learning, the authors train sub-classifiers on a subset of features instead of training one single classifier. The final classification decision was made by combining the decisions of the sub-classifiers. The ensemble learning method was demonstrated to be effective when the attack was made in black-box settings. When the attacker had information regarding the target model, there was a significant drop in defense performance.
\par 
Securedroid is a feature selection-based adversarial defense mechanism. Proposed by Chen et al. \cite{chen2017securedroid}, this method reduces the use of easily manipulated features in a classifier so that the adversary has to perturb many features to produce misclassification. The feature selection is made based on an `importance' metric given by $ I(i) \propto \frac{|\theta_i|}{c_i}$ where $\theta_i$ is the weight of a feature $i$ and $c_i$ is the manipulation cost for the feature modification determined empirically. Azmoodeh et al. \cite{azmoodeh2018robust} employed a class-wise feature selection method to remove unimportant Opcode features in malware classification. They employ Eigen-space analysis to extract important Opcode features to resist junk-code insertion attacks. Then, they create a graph of selected features and train a robust malware classifier.
\par 
Wang et al. \cite{wang2017adversary} propose a random feature nullification method for building adversary-resistant DNN. This method adds a stochastic layer between the input and the first hidden layer that randomly nullifies the input features in both the train and test phases. This layer ensures that the adversary does not have easy access to features for creating adversarial samples.  
\par 
Li et al. \cite{li2020adversarial} combine ensemble network with adversarial training technique. The hardened model was able to defend against a broad range of attacks but was vulnerable to mimicry attacks. In 2021, Li et al. \cite{li2021framework} again propose an ensemble framework for defense using a mixture of classifiers. Each sub-classifier is trained using adversarial training. All sub-classifiers are combined to form an ensemble classifier for adversarially robust classification.  This approach leverages the strengths of multiple models, potentially improving the robustness and generalizability of the defense.

\begin{table*}[]
\centering
\caption{Adversarial defenses on android malware classifier covered in this paper. Acronyms: R (improving the robustness of a model), D (detecting adversarial samples)}
\label{tab:advAttacks1}
\resizebox{0.88\textwidth}{!}{%
\begin{tabular}{lccl}
\hline
\multicolumn{1}{c}{\textbf{Defense}} & \textbf{Year} & \textbf{Goal} & \multicolumn{1}{c}{\textbf{High level summary}} \\ \hline
Smutz et al. \cite{smutz2016tree} & 2016 & D & Propose ensemble classifier mutual agreement analysis to identify uncertainty in classification \\
Demontis et al. \cite{demontis2017yes} & 2017 & R & Modify the Drebin classifier by introducing a regularization on feature weights of classifier \\
Grosse et al. \cite{grosse2017adversarial} & 2017 & R & Evaluate existing defense techniques like defensive distillation and adversarial training \\
Yang et al. \cite{yang2017malware} & 2017 & R, D & Evaluate different defenses like adversarial training, weight bounding, and variant detector \\
Grosse et al. \cite{grosse2017statistical}  & 2017 & D & Uses the distinguishability of adversarial examples with the normal training data distribution\\
Wang et al. \cite{wang2017adversary} & 2017 & R & Propose a random feature nullification method for building adversary-resistant DNN. \\
Chen et al.\cite{chen2017securedroid}  & 2017 & R & Train a new classifier by reducing the use of easily manipulated features \\
Li et al. \cite{li2018hashtran} & 2018 & D & Propose a detector framework combining hash function transformation and denoising auto-encoder \\
Chen et al. \cite{chen2018droideye} & 2018 & R & Uses  the concept of count featurization for adversarial defense \\
Azmoodeh el al. \cite{azmoodeh2018robust} & 2018 & R & Employ a class-wise feature selection method to remove unimportant features \\
Chen et al. \cite{chen2019android} & 2019 & R & Evaluate adversarial training and ensemble learning methods for adversarial defense \\
Li et al. \cite{li2020adversarial} & 2020 & R & Combine ensemble defense network with adversarial training \\
Li et al. \cite{li2021framework} & 2021 & R & Propose an ensemble framework for defense using a mixture of classifiers \\
Labaca et al. \cite{labaca2021realizable} & 2021 & R & Propose adversarial training based defense using universal problem-space attacks \\
Wang et al. \cite{wang2021exposing} & 2021 & R & Improve the detector by removing features from the training dataset sensitive to model prediction \\
Li et al. \cite{li2021robust} & 2021 & D & Employ similarity constraint to squeeze space for the presence of adversarial examples using VAE \\
Yang et al. \cite{yang2021cade} & 2021 & D & Learn a distance function to measure dissimilarity between samples using contrastive learning \\ 
Berger et al. \cite{berger2022mamadroid2} & 2022 & R & Designed MaMaDroid 2.0 by fusing static features and flow graphs of android apps\\
Li et al. \cite{li2023pad} & 2023 & R & Introduce a novel adversarial training approach called Principled Adversarial Malware Detection (PAD) \\
\hline
\end{tabular}%
}
\end{table*}

\par 
DroidEye by Chen et al. \cite{chen2018droideye} is an adversarial defense approach using the concept of count featurization. In count featurization, the categorical output is represented by the conditional probability of the output class given the frequency of the feature observed in that class. DroidEye transforms the binary-encoded feature of Android apps into continuous probabilities that encode additional knowledge from the dataset. The classifier is trained on the \textit{count-featurized} probabilities features. This approach can potentially make the model more robust to adversarial attacks by reducing the impact of individual features and making it harder for an adversary to manipulate the model predictions by altering a single feature.
\par 

Wang et al. \cite{wang2021exposing} improve the detector by removing features from the training dataset that are sensitive to the model prediction accuracy. Using Shapley values, they measure the contribution of features to the prediction accuracy and remove highly sensitive features. Then, they retrain the model with the new training set. This gives them a new model with improved robustness against adversarial attacks. By removing sensitive features, they reduce the potential attack surface for an adversary, making it harder to find successful attack strategies.
Berger et al. \cite{berger2022mamadroid2} analyzed the permission set of Android applications and observed that using the full permission set from the Android dataset only increased the feature size. The less frequent permissions do not affect the detection process. Based on these insights, they designed an enhanced version of the MaMaDroid classifier called MaMaDroid 2.0 by fusing static features and flow graphs of Android apps using a decision tree model. This approach can potentially improve the model’s robustness and interpretability by focusing on the most informative features and using a model structure (decision tree) that is easy to understand and interpret.
\par 
Labaca et al. \cite{labaca2021realizable} propose adversarial training-based defense using universal problem-space attacks and demonstrate that universal perturbations in the problem space capture adversarial vulnerabilities that harden a classifier more effectively with adversarial training. Li et al. \cite{li2023pad} introduce a novel adversarial training approach called Principled Adversarial Malware Detection (PAD), which guarantees convergence for robust optimization methods. PAD utilizes a learnable convex measurement to quantify distribution-wise discrete perturbations that safeguard malware detectors from adversarial attacks. To improve defense effectiveness, a new mixture of attacks is proposed to instantiate PAD.
\par 
\subsubsection{Adversarial detection} Smutz et al. \cite{smutz2016tree} propose an ensemble method called ensemble classifier mutual agreement analysis where the goal is to identify uncertainty in classification decisions by introspecting the decision given by individual classifiers in an ensemble. If each classifier in an ensemble has a similar prediction, the classifier decision is accurate. On the other hand, if most individual classifiers have conflicting predictions, the classifier returns uncertain results instead of benign or malicious predictions. 
\par 
Yang et al. \cite{yang2017malware} design a variant detector to identify if an application is a mutated version of an existing application. This variant detector is a classifier network trained with mutation features generated from each pair of app features that explain the feature difference between original and mutated apps. Grosse et al. \cite{grosse2017statistical} use the distinguishability of adversarial examples with the normal training data distribution for adversarial sample detection. They propose to detect adversarial samples as an outlier detection system. They train a model to recognize adversarial samples as outliers since adversarial examples are not representative of training data distribution and lie in unexpected regions of the model’s output surface.
\par 
Li et al.\cite{li2018hashtran} propose a detector framework combining hash function transformation and denoising auto-encoder. A hashing layer transforms the input features into a vector representation using a locality-sensitive hash function. A denoising auto-encoder receives this hashed vector as input that understands its locality information in the latent space. The auto-encoder learns the distribution of the hash by attempting to recreate the input. In the testing phase, the classifier rejects an adversarial sample as it produces a high reconstruction error. \par 
In \cite{li2021robust}, Li et al. employ similarity constraint in order to squeeze space for the presence of adversarial examples. They use a VAE to differentiate between benign and adversarial samples based on the reconstruction errors. They modify the loss function for VAE to disentangle features of different classes. This allows the VAE to learn a compact representation of the input classes in such a way that the distance between samples of the same class is small and the distance between samples of different classes is larger. Contrastive Auto-encoder for Drifting Detection and Explanation (CADE) \cite{yang2021cade} detects drifting samples and provides explanations for detected drift. Drifting samples are test data distribution shifting from training data similar to adversarial examples. CADE maps the data samples into a low dimensional space and learns a distance function to measure dissimilarity between samples using contrastive learning \cite{hadsell2006dimensionality}. A test sample is declared as out-of-distribution if the distance exceeds a threshold measure.

\section{Evaluation metrics}
In this section, we list out essential evaluation metrics that need to be computed to compare the performance of adversarial attacks and defenses. 
\subsection{Adversarial attack}

\textbf{Evasion robustness (ER):} It measures the robustness of the detection system against adversarial attacks and is measured by the true positive rate (TPR) of the malicious applications \cite{berger2022mamadroid2}. We compute ER as the ratio of malicious instances successfully detected to the total number of malicious applications (including adversarially modified malware).

\textbf{Evasion Increase Rate (EIR):} ER measures the true positive rate when adversarial samples are introduced. A more precise analysis of evasion attack compares the original true positive rate (without adversarial attack) with the true positive rate when an adversarial attack occurs \cite{chen2021fooling}. Let OR represent the original TPR. Then, EIR is given by Equation \ref{eqn:eir}. 
    \begin{equation}\label{eqn:eir}
        EIR = 1 - \frac{ER}{OR}
    \end{equation}

 \textbf{Defense Reciprocal Rank (DRR):} The Defense Reciprocal Rank (DRR) is based on the idea that a successful evasion attack is not solely determined by the correctness of the classification but also by the confidence of the classifier and proposed by Brama et al. \cite{brama2023evaluation}. 
    The probability assigned to the true class $i$ is denoted by $\hat{P}_i$, and the rank of that class within the ordered predictions list is represented by $R_i$. For the chosen class, $\hat{P}_i$ ranges between $0-1$, while for the second-best class, $\hat{P}_i$ ranges between $0-0.5$. The ranks for binary classification are 1 and 2, where 1 is the highest rank. 
    DRR is calculated using Equation \ref{eqn:drr} for a given sample x:
    \begin{equation}\label{eqn:drr}
        DRR(x) = \frac{\hat{P}_i}{R_i + 1} + \frac{1}{R_i + 1}
    \end{equation}

    The overall DRR for a classifier for a test set X is given by Equation \ref{eqn:testdrr}:
    \begin{equation}\label{eqn:testdrr}
        DRR(X) = \frac{\sum_{xinX} DRR(x)}{|X|}
    \end{equation}

\textbf{Model Reliability Assessment:} In binary classification, the model outputs a probability ($p$) for the target class and its complement probability ($\hat{p}$) for second class. Let the entropy of the target class and the second class for each sample be denoted as $s$, i.e., $s=E(p, \hat{p})$ where $E$ is the entropy function. To assess the reliability of the model, one can compute the Shannon Entropy of the probabilities as suggested by Nguyen \cite{nguyen2020ensemble}. The reliability is based on the complement of the average entropy of the test sample set, where a higher score is indicative of greater confidence by the classifier. The reliability is defined by Eq. \ref{eqn:modelreliability} with respect to the classifier, the set of test samples (X), and the set of entropy (S) for the test set X.
    \begin{equation}\label{eqn:modelreliability}
        Re = 1 - \frac{S}{|X|}
    \end{equation}

\subsection{Adversarial defense}
\textbf{Accuracy:} The evaluation of model robustness against adversarial attacks employs accuracy as a key metric. Firstly, normal accuracy measures the model's performance on clean, non-adversarial examples. A robust model should exhibit high accuracy on normal inputs, indicating its ability to correctly classify non-adversarial examples, remaining unaffected by adversarial perturbations. Secondly, adversarial accuracy assesses the model's ability to correctly classify inputs that have been modified by adversarial perturbations. It specifically evaluates the performance of adversarial examples. Additionally, the attack success rate is measured, which measures the percentage of adversarial examples misclassified by the model. A robust model should achieve a low attack success rate, indicating resilience against adversarial attacks.

\textbf{True Positive Rate (TPR):} The performance of adversarial sample detectors is evaluated by calculating the ratio of correctly identified adversarial samples (true positives) to the total number of adversarial samples. This metric provides insight into the ability of a detector to correctly identify adversarial samples, with a higher TPR indicating better performance.

\textbf{False Positive Rate (FPR):} In addition to measuring the detection rate of the detector (TPR), assessing its performance necessitates the computation of the false positive rate (FPR). The FPR quantifies the detector's susceptibility to incorrectly classifying normal examples as adversarial. A lower FPR indicates that the detector is less likely to misclassify normal examples as adversarial, which is critical for maintaining the usability of the system.

\textbf{AU-ROC:} AUROC is the probability that a positive test instance has a higher detector score than a negative example and is regarded as a threshold-independent measure of detector performance \cite{davis2006relationship}. This metric provides a quantitative measure of the performance of a detector across all possible classification thresholds, or for comparing different detectors or tuning the performance of a single detector.

\section{Guidelines for building robust android malware classifiers}\label{sec:guidelines}
We now propose guidelines to consider for designing robust defenses against evasion attacks on Android malware classifiers:
\begin{enumerate}
    \item \emph{Build accurate threat model}: Defenders should accurately model the threat attacks to design suitable defenses. Threat modeling principles are explained in Section \ref{threatModel}. Other frameworks include FAIL\cite{suciu2018does} that can help in avoiding any unrealistic assumptions. Understanding the potential threats and how they operate can help in developing more effective defenses.

    \item \emph{Perform strong attacks against the defense}: Adversarial defenses must withstand the strongest attacks. Optimization-based attacks like C\&W \cite{carlini2017towards}, and Madry \cite{madry2018towards} are the most powerful attacks so far \cite{carlini2019evaluating}. Defense against weaker attacks like FGSM \cite{goodfellow2015explaining} does not prove improved robustness \cite{carlini2017adversarial}. Therefore, a defender should test their defense against white-box and black-box attacks of both targeted and un-targeted attack types. This rigorous testing can help identify potential weaknesses in the defense strategy.

    \item \emph{Inspect if the attack is transferable}: The transferability property of adversarial examples is one of the primary reasons behind the vulnerability of many defenses \cite{tramer2017space}. Therefore, defenders should ensure that a defense algorithm breaks this property of adversarial attack. This involves testing the defense with different models and configurations to ensure its effectiveness across various scenarios.

    \item \emph{Evaluate security evaluation curve}: Defenders should also evaluate the defense strategy with an increasing attack strength. This information is plotted as a security evaluation curve that shows how the defense algorithm performs under increasing strength of the attack \cite{biggio2018wild}. Relying solely on the assumption of small perturbations for the attack does not capture the full range of attacks encountered in real-life scenarios. This evaluation can provide valuable insights into the defense's performance and help identify areas for improvement.
\end{enumerate}

\section{Research directions}\label{sec:research}

\textbf{Adversarial machine learning and interpretability:} 
Machine learning classifiers, when subjected to adversarial attacks, often make erroneous decisions with very high confidence. This is a significant issue, especially considering that these attacks are entirely unknown forms. Consequently, building robust and resilient models to withstand such attacks is a challenging task. The main question thus lies in understanding why classifiers trained to generalize well to unseen data succumb to small perturbations. Only by understanding the root cause can we devise effective solutions. Future research, hence, should focus on understanding the paradigm of adversarial vulnerability. Gathering theoretical or empirical evidence on why machine learning algorithms behave unexpectedly to adversarial attacks will provide essential design techniques for defenders. In this regard, exploring the intersection of adversarial learning and machine learning explainability can provide compelling insights \cite{zheng2020understanding}. This could involve developing new methods for visualizing and interpreting adversarial attacks or devising novel metrics for quantifying the interpretability of a model under attack.
\par
\textbf{Quantifying robustness:}  Despite motivated efforts in quantifying the robustness of a classifier \cite{carlini2017provably}, no approaches are universally acceptable. A mathematical bound on the robustness of classifiers depending on certain architectural or dataset conditions and constraints can help understand the model's confidence in adversarial settings. New attacks will constantly break heuristics-based defenses. Hence, a solid mathematical understanding of adversarial robustness is necessary to design resilient classifiers. Future research could focus on developing more rigorous and universally acceptable measures of robustness or exploring how different architectural choices or training strategies impact a model’s robustness to adversarial attacks. Additionally, investigating how robustness measures correlate with other desirable properties of machine learning models, such as accuracy or fairness, can also be explored.

\par
\textbf{Robust feature selection:} Recent research has presented adversarial property as an inherent quality of the dataset, shedding light on the vulnerability of machine learning models to adversarial attacks. To address this challenge, it is essential to delve deeper into the identification and selection of robust features. This process involves analyzing the dataset and identifying features that are less susceptible to adversarial manipulations. Robust feature selection identifies discriminative features that exhibit consistent patterns across both normal and adversarial examples. By incorporating robust features into the model's decision-making process, we can enhance the model's ability to withstand adversarial perturbations. This potentially involves developing new algorithms for robust feature selection or exploring how different types of features contribute to the robustness of a model against adversarial attacks. Additionally, devising methods to automatically select robust features during the training process, thereby improving the model’s resilience without requiring manual intervention, could be explored.
\par 

\section{Conclusion}
In this survey, we explored fundamental research on adversarial attacks and defenses on Android malware classifiers and presented the central hypothesis on adversarial vulnerability. We summarized evasion attacks and defenses, highlighting the key techniques and strategies used in each case. We elucidated the metrics required for adversarial machine learning, providing a comprehensive overview of the measures used to evaluate the performance and robustness of models in adversarial settings. We also formulated our study into guidelines to assist in defense designs, offering practical advice and recommendations based on our findings. Finally, we presented future research directions in the field, identifying key areas where further investigation could lead to significant advancements in the understanding and mitigation of adversarial attacks.

\bibliographystyle{ACM-Reference-Format}
\bibliography{main}

\end{document}